\documentclass[aps,prb,twocolumn,floatfix,amsmath,amssymb,superscriptaddress]{revtex4}
\usepackage[final]{graphicx}
\usepackage{bm}
\usepackage{afterpage}
\usepackage{color}
\usepackage{comment}
\usepackage[colorlinks=true,urlcolor=blue,linkcolor=blue,citecolor=blue]{hyperref}

\begin{document}

\title{Exact Demonstration of pair-density-wave superconductivity in the $\sigma_z$-Hubbard model}

\author{Xingchuan Zhu\footnote{These authors contributed equally to this work} }

\affiliation{Interdisciplinary Center for Fundamental and Frontier Sciences, Nanjing University of Science and Technology, Jiangyin, Jiangsu 214443, P. R. China}

\author{Junsong Sun$^*$}
\affiliation{School of Physics, Beihang University,
Beijing, 100191, China}

\author{Shou-Shu Gong}
\affiliation{School of Physical Sciences, Great Bay University, Dongguan 523000, China, and \\
Great Bay Institute for Advanced Study, Dongguan 523000, China}

\author{Wen Huang}
\affiliation{Shenzhen Institute for Quantum Science and Engineering,
Southern University of Science and Technology, Shenzhen 518055, Guangdong, China}

\author{Shiping Feng}
\affiliation{ Department of Physics,  Beijing Normal University, Beijing, 100875, China}

\author{Richard T. Scalettar}
\affiliation{Department of Physics and Astronomy, University of California, Davis, CA 95616, USA}

\author{Huaiming Guo}
\email{hmguo@buaa.edu.cn}
\affiliation{School of Physics, Beihang University,
Beijing, 100191, China}

\begin{abstract}
Describing and achieving `unconventional' superconductivity remains a forefront challenge in quantum many-body physics.
Here we use a unitary mapping, combined with the well-established 
properties of the attractive Hubbard model to demonstrate rigorously
a Hamiltonian with a low temperature pair-density-wave (PDW) phase.
We also show that the same mapping, when applied to the widely
accepted properties of the repulsive Hubbard model, leads to
a Hamiltonian exhibiting triplet $d$-wave PDW superconductivity and an unusual
combination of ferro- and antiferro-magnetic spin correlations.
We then demonstrate the persistence of the $d$-wave PDW in a Hamiltonian derived from the mapping of the extended $t$-$J$ model in the large-$U$ limit. Furthermore, through strategic manipulation of the nearest-neighbor hopping signs of spin-down electrons, we illustrate the attainability of PDW superconductivity at other momenta. The intertwining of different magnetic and exotic pairing correlations
noted here may have connections to experimental observations in spin-triplet candidates like UTe$_2$.
\end{abstract}


\maketitle
\textit{Introduction.---}
Conventional BCS superconductivity describes the pairing of a singlet
($s$-wave) pair of fermions with zero momentum, that is, a non-zero expectation value of the off diagonal order parameter $\Delta_s(k) = c_{-k\downarrow}^{\phantom{\dagger}} c_{k\uparrow}^{\phantom{\dagger}} $.
This low temperature phase is typically achieved through an effective
retarded attractive interaction mediated by electron-phonon coupling.
Shortly after BCS theory, the possibility of non-zero momentum pairs
was noted by Fulde and Ferrell~\cite{fulde1964superconductivity}, 
and by Larkin and Ovchinnikov~\cite{osti_4653415}.
Achieving such `FFLO' pairing proved very challenging but has been reported in heavy fermion systems such as CeCoIn$_5$~\cite{bianchi2003possible,matsuda2007fulde}, and in ultracold atoms~\cite{partridge2006pairing,parish2007quasi,kinnunen2018fulde}.

Other `unconventional' superconductors include
those in which the Cooper wavefunction exhibits more complex patterns in real
space, such as the nodes in the $d$-wave symmetry pairs
of the cuprate materials~ \cite{RevModPhys.72.969,RevModPhys.75.473,RevModPhys.78.17},
triplet superconductors which have nonzero total spin~\cite{RevModPhys.75.657},
pair density waves~\cite{agterberg2020physics} in which the order parameter varies 
spatially with vanishing spatial average, and $\eta$-pairs which are exact 
(high energy) 
eigenstates of the Hubbard Hamiltonian exhibiting off diagonal long range order~\cite{yang1989eta,singh1991exact,chen2011SO4},
{\it etc}.

Some of the unconventional superconductors noted above are (easily) achieved 
experimentally; others are less so.  A key question for
theory is what Hamiltonians give rise to the different types of `exotic' pairing.  
For example, in the case of the cuprates, the sufficiency of the repulsive Hubbard model continues to be debated~\cite{arovas2022hubbard,qin2022hubbard}.
As for a PDW phase, obtaining a stable one in two dimensions is even a greater challenge~\cite{PhysRevLett.88.117001,PhysRevLett.113.046402,Huang2021111010,PhysRevB.81.020511,PhysRevX.4.031017,Berg_2009,PhysRevB.96.224503,PhysRevB.108.174506,PhysRevLett.130.026001,Jiang2023Aug}.
In this paper, we present a pathway towards realizing PDW superconductivity in the
$\sigma_z$-Hubbard model. Our key observation is that 
a unitary transformation
combined with known results of the conventional Hubbard model, allows us to identify Hamiltonians
which rigorously must exhibit low temperature unconventional PDW superconductivity.

\textit{The Hubbard model.---}
We begin with the celebrated Hubbard model
\begin{align}
{\cal H} = -&t \sum_{\langle ij \rangle \sigma}
\big( c_{i\sigma}^{\dagger}  c_{j\sigma}^{\phantom{\dagger}} + 
c_{j\sigma}^{\dagger}  c_{i\sigma}^{\phantom{\dagger}} \big) 
- \mu \sum_{i} \big(n_{i\uparrow}^{\phantom{\dagger}} + n_{i\downarrow}^{\phantom{\dagger}}\big)
\nonumber \\
+ &U \sum_{i} \big(n_{i\uparrow}^{\phantom{\dagger}}-\frac{1}{2}\big)  \,
\big( n_{i\downarrow}^{\phantom{\dagger}}-\frac{1}{2} \big)
\end{align}
which describes spin $\sigma = \uparrow,\downarrow$ fermions hopping on a lattice
and interacting with an on-site interaction $U$.
When the interaction is attractive ($U<0$)
the phase diagram on a square lattice is well-understood qualitatively and quantitatively~\cite{PhysRevLett.66.946,PhysRevLett.62.1407,PhysRevB.69.184501}: 
At half-filling ($\mu=0$) the ground state exhibits simultaneous long range charge density wave 
(CDW)
and $s$-wave superconducting (SC) orders.  When doped ($\mu \neq 0$) the SC-CDW
degeneracy is broken, and there is a finite temperature (Kosterlitz-Thouless)
transition to a SC phase.
This description has been confirmed by Quantum Monte Carlo (QMC) calculations 
which, 
owing to the absence of a sign problem, can be carried out to arbitrarily low temperatures.

Full understanding of the repulsive model with a large $U > 0$ is more elusive.  
At half-filling
there is long range antiferromagnetic (AF) order which occurs only at $T=0$
on a square lattice owing to the continuous Heisenberg spin symmetry
and the Mermin-Wagner theorem~\cite{mermin1966absence}.  However, when doped, QMC fails to
reach low $T$ because of the sign problem.  
A $d$-wave SC phase, with intricate
`striped' charge and spin patterns is suggested by many calculations [sometimes with the addition of further next-near-neighbor (NNN) hopping], but the final
determination of the various orders remains under discussion~\cite{PhysRevX.5.041041,huang2017numerical,zheng2017stripe,Jiang2019Sep,Jiang_PRR_2020,Chung_PRB_2020,Qin_PRX_2020,Xu2022,Xu2023,jiang2024}.

Before introducing the main results of this work,
it is useful to review the well-known 
(partial) particle-hole transformation 
$c_{i \downarrow}^{\phantom{\dagger}}
\rightarrow (-1)^{i_x+i_y} \, c_{i \downarrow}^{\dagger}$ 
which links the descriptions of the properties of the attractive and repulsive cases.
Here $(-1)^{i_x+i_y}$ indicates opposite phases
on the two sublattices of the (bipartite) square lattice.
Under this transformation, 
the kinetic energy remains unchanged.
The down spin density
$n_{i \downarrow}^{\phantom{\dagger}}
\leftrightarrow
1 - n_{i \downarrow}^{\phantom{\dagger}}$
and, as a consequence  the sign of $U$ is reversed, mapping attraction to repulsion
and vice-versa.  The roles of charge and spin operators are interchanged
$n_{i\uparrow}^{\phantom{\dagger}} + n_{i\downarrow}^{\phantom{\dagger}}
\leftrightarrow
n_{i\uparrow}^{\phantom{\dagger}} - n_{i\downarrow}^{\phantom{\dagger}}$,
so that chemical potential $\mu$ and Zeeman $B_z$ terms map into one another
(to within an irrelevant energy shift) and
correlations of the $Z$ component of spin map onto density correlations.
Finally, the $XY$ spin operators map onto $s$-wave pairing
$c_{i\uparrow}^{\dagger} c_{i\downarrow}^{\phantom{\dagger}} 
\leftrightarrow
c_{i\uparrow}^{\dagger} c_{i\downarrow}^{\dagger} $.

With those mappings in place, the connections between the attractive and repulsive Hubbard models
become clear.  The fact that the square lattice
repulsive Hubbard model has degenerate $Z$ and $XY$ spin
order in its ground state and half-filling immediately implies the degenerate
CDW and SC patterns in the attractive case.  Likewise, the fact that a Zeeman
field $B_z$ causes AF Heisenberg spins to `lie down' and order in the $XY$ 
plane perpendicular to
the field is then connected to the preference for SC correlations over CDW ones in the
attractive Hubbard model for $\mu$ nonzero. We will now show how an alternate canonical transformation
lends similar insight into exotic superconductivity.

\begin{figure}[htbp]
\centering \includegraphics[width=8.0cm]{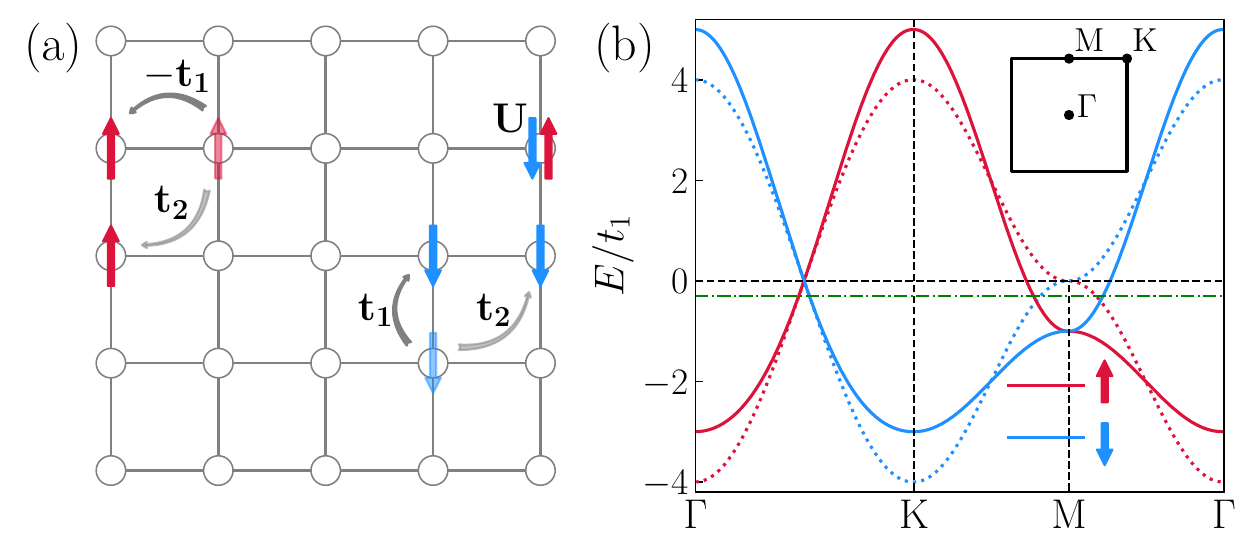} \caption{(a) A schematic view of the extended $\sigma_z$-Hubbard model on a square lattice, where $t_1$($-t_1$) is the NN hopping parameter of the spin-up (spin-down) fermion, $t_2$ represents the NNN hopping amplitude and $U$ is the on-site Hubbard interaction. Up and down arrows correspond to spin-up and spin-down electrons, respectively. (b) Band structures of the non-interacting terms in Eq.(1) with $t_2=0$ (dotted line) and $t_2=0.25$ (solid line). Inset displays the first Brillouin zone, on which the high-symmetry points are marked. }
\label{fig1}
\end{figure}

\textit{Attractive $\sigma_z$-Hubbard model.---}
We then apply the unitary transformation, $c_{i\downarrow}\rightarrow \text{sgn}(i)c_{i\downarrow}$,
to the attractive Hubbard model, resulting in the $\sigma_z$-Hubbard model defined
by the Hamiltonian~\cite{higher-order-kagome,otsuka2021higher,PhysRevB.105.245131},
\begin{align}\label{eq1}
{\cal H}_{\sigma_z}=-&t \sum_{\langle i j\rangle} \sum_{\alpha \beta}c_{i \alpha}^{\dagger} \sigma^{\alpha \beta}_z c_{j \beta}-\mu \sum_{i,\alpha}n_{i\alpha} \\ \nonumber
+&U\sum_{i}(n_{i\uparrow}-\frac{1}{2})(n_{i\downarrow}-\frac{1}{2})
\end{align}
where $\sigma_z$ represents the $Z$-component of the Pauli matrix, resulting in opposite signs in the hopping amplitudes for the spin-up and spin-down subsystems [see Fig.~\ref{fig1}(a)].

The phase diagram of the attractive Hubbard model can be transformed back to derive that of the attractive $\sigma_z$-Hubbard model. It is observed that while the CDW remains unaffected, the $s$-wave SC phase is altered. Specifically, the on-site pairing transforms back as $\Delta_j= c_{j \downarrow} c_{j \uparrow} \rightarrow \Delta_j=\text{sgn}(j)c_{j \downarrow} c_{j \uparrow} $. Therefore, the pairing remains on-site but with an alternating sign, indicating that the system displays $s$-wave PDW superconductivity. The pairing function can be written as $\Delta^{\dagger}=\frac{1}{\sqrt{N}} \sum_j (-1)^{j_x+j_y} c_{j}^{\dagger} c_{j}^{\dagger}=\frac{1}{\sqrt{N}} \sum_{\bf k} c_{\bf k}^{\dagger} c_{{\bf -k}+{\bf K}_0}^{\dagger}$ with ${\bf K}_0=(\pm\pi,\pm\pi)$. 
Therefore a PDW state, in which an electron at momentum ${\bf k}$
pairs up with another at momentum $-{\bf k}+{\bf K}_0$, resulting
in a Cooper pair carrying net momentum ${\bf K}_0$, must 
rigorously be the low temperature phase of the attractive $\sigma_z$-Hubbard model.

\begin{figure}[htbp]
\centering \includegraphics[width=8.0cm]{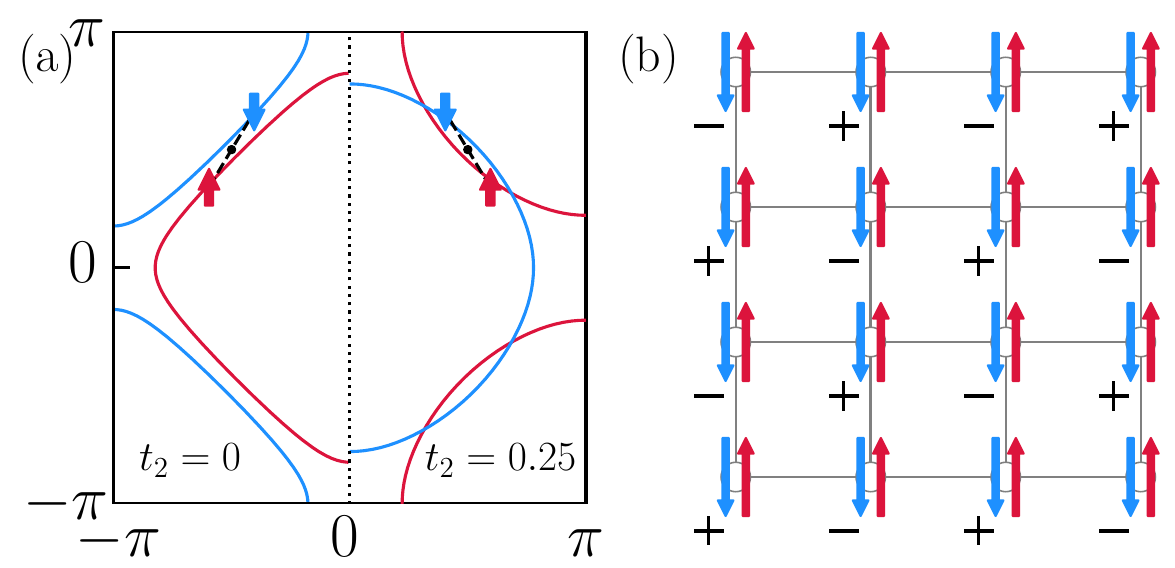} \caption{(a) Fermi surfaces of the spin-up and spin-down
electrons at $\mu=-0.3$, in which a pair of electrons with different spins on the nested Fermi surface is demonstrated. The Fermi surfaces are $C_4$ symmetric,  therefore only left (right) half of them are shown for $t_2=0$ $(t_2=0.25)$. (b) A schematic view of the $s$-wave PDW with center of mass momentum $(\pi,\pi)$ in the attractive $\sigma_z$-Hubbard model on a square lattice. }
\label{fig2}
\end{figure}

It has been well established that the Fermi surface topology plays a crucial role in Cooper pair formation. To investigate the origin of PDW in the $\sigma_z$-Hubbard model, we plot the non-interacting Fermi surface in Fig.~\ref{fig2}(a). In contrast to the normal spin-independent hopping scenario where the Fermi surface is spin-degenerate, the $\sigma_z$ hopping term generates a spin-dependent Fermi surface. Since the dispersion of the two spin species satisfies the condition $\xi_{\uparrow,\bf k}$ = $\xi_{\downarrow,-\bf k+\bf K_0}$, the spin-up and spin-down Fermi surfaces are of identical shape and are centered around the $\Gamma$ and $K$ points, respectively. The above relation also indicates perfect nesting in the particle-particle channel with center of mass momentum ${\bf K}_0$. Hence, in the presence of on-site attractive interaction, a PDW order with modulation wavevector ${\bf K}_0$ will develop.

\textit{Repulsive $\sigma_z$-Hubbard model.---}
We next consider the repulsive case, $U>0$. Introducing a NNN hopping term $t_2$ into the Hamiltonian in Eq.~\eqref{eq1}, the total Hamiltonian becomes:
\begin{align}\label{eq3}
{\cal H}=&-t_1 \sum_{\langle i j\rangle} \sum_{\alpha \beta}c_{i \alpha}^{\dagger} \sigma^{\alpha \beta}_z c_{j \beta}+t_2\sum_{\langle\langle i j\rangle\rangle \alpha} \hat{c}_{i \alpha}^{+} \hat{c}_{j \alpha} \\ \nonumber
&+U\sum_{i}(n_{i\uparrow}-\frac{1}{2})(n_{i\downarrow}-\frac{1}{2})-\mu \sum_{i,\alpha}n_{i\alpha},
\end{align}
where $\langle\langle ij\rangle\rangle$ denotes next-nearest neighbors, and $t_2$ is the NNN hopping amplitude.
Under the same unitary transformation in the previous section, the above extended $\sigma_z$-Hubbard model transforms into a normal extended one. 
The inclusion of the NNN hopping term is essential at large $U$ as it may play a key role in generating long-range SC correlation and establishing a delicate balance between CDW, spin density wave, and superconductivity~\cite{xiao2023temperature,mai2023robust,Jiang_PRR_2020,jiang2024}.
Recent comprehensive density matrix renormalization group (DMRG) studies have revealed the intertwined CDW and SC correlations on 4-leg systems~\cite{Jiang2019Sep,Jiang_PRR_2020,Chung_PRB_2020}, as well as an emergent $d$-wave SC phase on wider systems with a moderate $t_2 > 0$~\cite{jiang2024}.

The magnetic order can be characterized by the spin correlation functions defined as \(C^{z}(ij) = \langle S^z_iS^z_{j} \rangle = \langle (n_{i\uparrow}-n_{i\downarrow})(n_{j\uparrow}-n_{j\downarrow}) \rangle\) and \(C^{xy}(ij) = \langle S^{+}_iS^{-}_{j} \rangle = \langle c_{i \uparrow}^{+} c_{i \downarrow} c_{j \downarrow}^{+} c_{j \uparrow} \rangle\). Under the transformation, \(C^{xy}(ij)\) will change its sign when \(i\) and \(j\) belong to different sublattices, whereas \(C^{z}(ij)\) will remain unchanged. The magnetic properties of the normal extended Hubbard model [connected to Eq.~\eqref{eq3} through the unitary transformation] have been well established at half filling, exhibiting an AF ground state\cite{PhysRevB.80.075116,zheng2017stripe,arovas2022hubbard,qin2022hubbard}. Consequently, the Hamiltonian in Eq.~\eqref{eq3} will exhibit unconventional long-range magnetic order, which is ferromagnetic (antiferromagnetic) in the \(XY\) plane (\(Z\) direction).

Similarly, in diagnosing the SC order, if the pair correlation function involves the spin-singlet pair annihilation operator \(\Delta(ij) = \frac{1}{\sqrt{2}}(c_{i \uparrow} c_{j \downarrow} - c_{i \downarrow} c_{j \uparrow})\), it is straightforward to infer that the transformation will convert the spin-singlet pair operator to \(\Delta(ij) = \pm\frac{1}{\sqrt{2}}(c_{i \uparrow} c_{j \downarrow} + c_{i \downarrow} c_{j \uparrow})\) for the nearest-neighbor (NN) pairing [as is the case for the most-studied $d$-wave in the normal repulsive Hubbard model], where $+(-)$ corresponds to the negative sign of the transformation situated on site $i(j)$. Then, the NN pairings $c_{i\uparrow}c_{j\downarrow}$ and $c_{j\uparrow}c_{i\downarrow}$, which are equivalent in the spin-singlet scenario, will exhibit a sign difference under the unitary transformation, giving rise to a spin-triplet state. Therefore, the additional symbol \(\pm\) will not only produce a net momentum ${\bf K}_0$ for the $d$-wave pairs but also alter the nature of the pairing, leading to the emergence of a $d$-wave PDW triplet superconductor with a center of mass momentum ${\bf K}_0$~\cite{chen2020superconductivity}.

The various types of correlations mentioned above can be transformed back by the gauge transformation and are used to characterize the ground-state properties of the extended $\sigma_z$-Hubbard model. While the charge density correlations and spin-$z$ correlations remain unchanged, the SC correlations transform into those of spin-triplet $d$-wave pairings at momentum ${\bf K}_0$, and the transverse spin correlation transitions to be ferromagnetic (FM).
Therefore, considering that spin-singlet $d$-wave superconductivity dominates in the doped normal extended Hubbard model~\cite{Jiang_PRR_2020,Chung_PRB_2020,Xu2023,jiang2024}, it is reasonable to suggest that the extended $\sigma_z$-Hubbard model may support a spin-triplet $d$-wave PDW SC ground state with the center of mass momentum ${\bf K}_0$. It is noted that the conversion of the pairing symmetry is accompanied by changes in the transverse magnetic property, i.e., from AF to FM corresponding to the shift from singlet to triplet pairings. This may imply the significant role of FM spin fluctuations in mediating the formation of spin-triplet pairs of electrons.

\begin{figure}[htbp]
\centering \includegraphics[width=8.0cm]{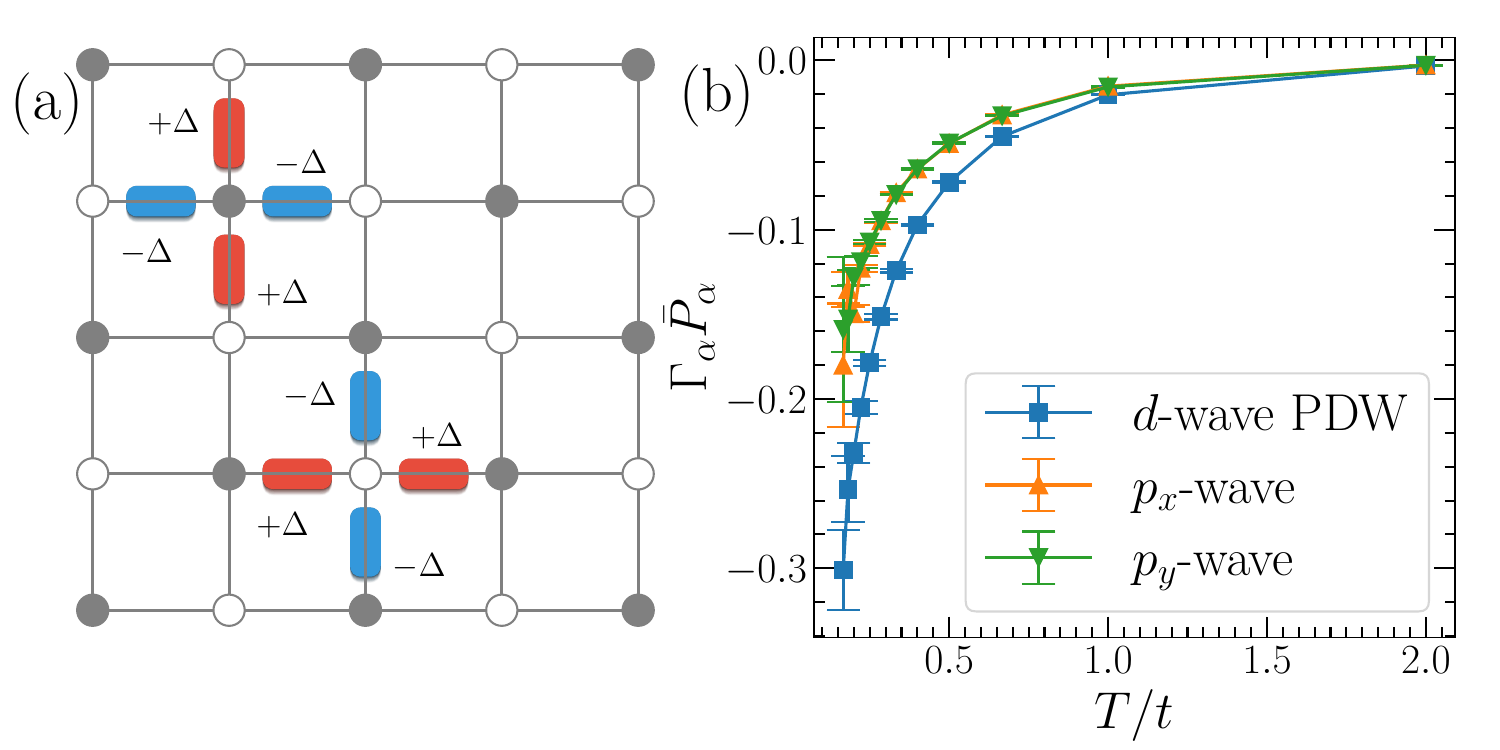} \caption{(a) A schematic demonstration of the $d$-wave PDW with center of mass momentum $(\pi, \pi)$ in the extended repulisive $\sigma_z$-Hubbard model on a square lattice. (b) The measured $\Gamma_\alpha \bar{P}_\alpha$ of the pairing instability as a function of temperature for various paring symmetries at a filling of $\rho=0.95$. The parameters used are $t'/t=0.25$ and $U/t=4$, with a lattice size of $L=8$.}
\label{fig3}
\end{figure}

To confirm the existence of the $d$-wave triplet SC at momentum ${\bf K}_0$ in the extended $\sigma_z$-Hubbard model given by Eq.~\eqref{eq3}, we conduct DQMC calculations of the $d$-wave pairing susceptibility at momentum ${\bf K}_0$, defined as follows~\cite{PhysRevB.39.839}:
$$
P_{d}=\frac{1}{N} \int_0^\beta d \tau \sum_{i j}\left\langle\Delta_i^d(\tau) \Delta_j^{d \dagger}(0)\right\rangle e^{i{\bf K}_0\cdot ({\bf r}_j-{\bf r}_i)},
$$
where $\Delta_i^d(\tau)=\sum_j f_{i j}^d e^{\tau H} c_{i \uparrow} c_{j \downarrow} e^{-\tau H}$ represents the time-dependent pairing operator with a form-factor $f_{i j}^d= 1(-1)$ for the bond in the $X(Y)$ direction between sites $i$ and $j$.
The interaction vertex $\Gamma_{d}$ can be extracted from $P_{d}$ and the uncorrelated susceptibility $\bar{P}_{d}$ as follows:
$\Gamma_{d}=\frac{1}{P_{d}}-\frac{1}{\bar{P}_{d}}$~\cite{PhysRevB.86.184506,PhysRevB.90.075121}. When $\Gamma_{d} \bar{P}_{d}<0$, the corresponding pairing interaction is attractive. As $\Gamma_{d} \bar{P}_{d} \rightarrow-1$, $P_{d}$ tends to diverge, indicating a SC instability.
Figure~\ref{fig3}(b) illustrates the product $\Gamma_{d} \bar{P}_{d}$ for the $d$-wave pairing susceptibility at momentum ${\bf K}_0$. As the temperature decreases, we find $\Gamma_{d} \bar{P}_{d}$ is the most negative, suggesting this pairing will dominate the SC instability. For comparison, we also calculate $\Gamma_{\alpha} \bar{P}_{\alpha}$ for the $p$-wave triplet at zero momentum, which is less dominant.

\textit{The large-$U$ limit.---}
In the large-$U$ limit, the double occupancy on a lattice site is excluded, and the extended $\sigma_z$-Hubbard model is reduced to an extended $\sigma_z$-$t$-$J$-like Hamiltonian~\cite{charles1976},
\begin{align}\label{eq4}
{\cal H}=&-t_1 \sum_{\langle i j\rangle} \sum_{\alpha \beta}c_{i \alpha}^{\dagger} \sigma^{\alpha \beta}_z c_{j \beta}+t_2\sum_{\langle\langle i j\rangle\rangle \alpha} \hat{c}_{i \alpha}^{+} \hat{c}_{j \alpha} \\ \nonumber
& -\sum_{\langle i j\rangle}[J_1\left(S^{x}_iS^{x}_j+S^{y}_iS^{y}_j\right)-J_1S^{z}_iS^{z}_j+J_1\frac{1}{4} \hat{n}_i \hat{n}_j] \\ \nonumber
& +\sum_{\langle\langle i j\rangle\rangle}J_2(\hat{\mathbf{S}}_i \cdot \hat{\mathbf{S}}_j-\frac{1}{4} \hat{n}_i \hat{n}_j),
\end{align}
where the exchange coupling is $J_{1(2)}=\frac{4 t_{1(2)}^2}{U}$ with the ratio $J_2/J_1=t^2_2/t^2_1$. Here, the FM nature of the $XY$ component of the NN Heisenberg term $J_1$ aligns with the magnetic properties observed in Eq.~\eqref{eq3}. Under the transformation, the above Hamiltonian becomes the normal extended $t$-$J$ model, on which
recent DMRG calculations have been conducted for six- and eight-leg cylinders, uncovering a robust $d$-wave SC phase in the case of electron doping ($t_2 > 0$)~\cite{Gong2021Aug,Jiang2021Aug,Jiang2021Oct,jiang_PRB_2023,Lu2024Feb,chen2023}. It has been demonstrated that the SC phase exhibits a power-law pairing correlation that decays much slower than the charge density and spin correlations. 
Furthermore, it is found that spin-singlet $d$-wave superconductivity can also emerge at the hole-doped side ($t_2 < 0$) near the optimal $1/8$ doping level on the wider $8$-leg system~\cite{Lu2024Feb,chen2023}. Correspondingly, it is reasonable to propose that the $\sigma_z$-$t$-$J$-like model in Eq.~\eqref{eq4} could potentially give rise to a $d$-wave PDW triplet superconductor with a center of mass
momentum ${\bf K}_0$ within the proper parameter region of the normal extended $t$-$J$ model, where the SC phase is observed~\cite{Gong2021Aug,Jiang2021Aug,Jiang2021Oct,jiang_PRB_2023,Lu2024Feb,chen2023,Qu2022Nov,Lu2023Mar,Jiang2022Nov}.

\begin{figure}[htbp]
\centering \includegraphics[width=8.0cm]{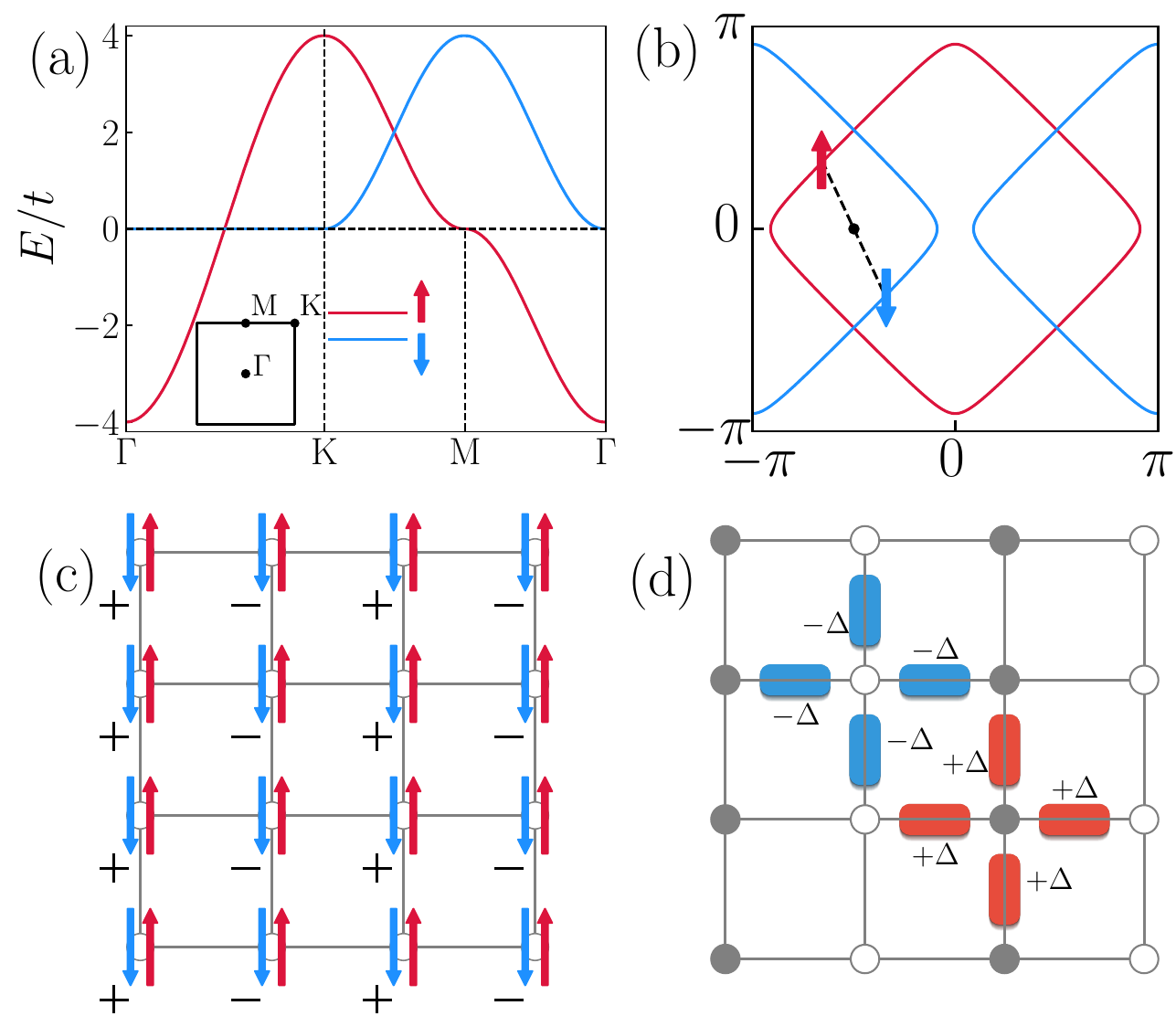} \caption{(a) Band structures of the
non-interacting Hamiltonian in Eq.~\eqref{eq5}. (b) Fermi surfaces of the spin-up and spin-down
electrons at filling $\rho=0.95$, in which the nesting of the Fermi surface is illustrated. Schematic views of the $s$-wave (c) and $d$-wave (d) PDWs with center of mass momentum $(\pi,0)$ in the modified $\sigma_z$-Hubbard model on a square lattice. }
\label{fig4}
\end{figure}

\textit{The $(\pi,0)$ PDW superconductivity.---}
The PDW SC ground state with a different center of mass momentum can be achieved by appropriately manipulating the NN hopping signs of the spin-down electrons. In the case of $(\pi,0)$, we can select the NN hoppings as follows:
\begin{align}\label{eq5}
{\cal H}_{(\pi,0)}=-t \sum_{i,\alpha \beta}c_{i \alpha}^{\dagger} \sigma^{\alpha \beta}_z c_{i\pm \hat{x} \beta}-t \sum_{i,\sigma}c_{i \sigma}^{\dagger}c_{i\pm \hat{y} \sigma},
\end{align}
where an additional sign is present when the spin-down electrons hop in the $x$ direction.
This additional sign can be eliminated by the following unitary transformation:
\begin{align}\label{eq6}
c_{i\downarrow}\rightarrow (-1)^{i_x}c_{i\downarrow}.
\end{align}
Through a similar analysis, the corresponding attractive Hubbard model supports a $s$-wave PDW state with a center of mass momentum of $(\pi,0)$.
Similarly, the dispersion of the two spins has the relation $\xi_{\uparrow,\bf k}$ = $\xi_{\downarrow,-\bf k+\bf K_0}$ with $\bf K_0=(\pi,0)$ and their Fermi surfaces are again nested in the particle-particle channel with center of mass momentum $\bf K_0$ [see Fig.~\ref{fig4}(b)].
By substituting the NN hopping term with the one in Eq.~\eqref{eq5} in the Hamiltonian Eq.~\eqref{eq3}, the modified repulsive Hubbard model exhibits an extended $s$-wave PDW state with a center of mass momentum of $(\pi,0)$. The pairings in the $X$ direction transition to triplet, accompanied by FM spin correlations in the $X$ component along this direction. These results extend to the corresponding model in the large-$U$ limit, which deviates from Eq.~\eqref{eq4} in the NN hoppings [replaced by Eq.~\eqref{eq5}] and NN exchange couplings (FM for the $X$ component along the $X$ direction). Finally, by rotating the Hamiltonian by $90$ degrees, PDW superconductivity with a  center of mass momentum $(0,\pi)$ can also be realized.

\textit{Conclusions.---}
We explicitly demonstrate the presence of PDW superconductivity in the $\sigma_z$-Hubbard model and its related extensions, such as incorporating long-range hoppings and the large-$U$ limit. These modified Hubbard or $t$-$J$ models can be transformed into normal ones through unitary transformations, where their physical properties have been thoroughly established. The attractive $\sigma_z$-Hubbard model supports an $s$-wave PDW phase, whereas the repulsive extended $\sigma_z$-Hubbard model features a $d$-wave PDW state. Both PDW phases possess a center-of-mass momentum at $(\pi,\pi)$. Specifically, the $d$-wave PDW at momentum $(\pi,\pi)$ is triplet, corresponding to which the spin correlations in the $XY$ component are ferromagnetic. The $d$-wave PDW persists in the extended $\sigma_z$-$t$-$J$-like model derived from the extended $\sigma_z$-Hubbard model in the large-$U$ limit.
Finally, we discover that a PDW superconductivity at momenutm $(\pi,0)$ can also be achieved by appropriately manipulating the NN hopping signs of the spin-down electrons.
Our study provides a microscopic mechanism for the PDW superconductivity, and will deepen the understanding of this exotic SC state~\cite{aoki2019review,PhysRevLett.60.615,RevModPhys.74.235,chen2021roton,gu2023detection,duan2021resonance}.
Specifically, while recent experiments have identified UTe$_2$ as a candidate for a spin-triplet PDW state near an FM instability\cite{gu2023detection}, the AF fluctuations detected by inelastic neutron scattering seem highly unusual\cite{duan2021resonance}. Nonetheless, the $d$-wave PDW at momentum $(\pi,\pi)$ mentioned here inherently exhibits a coexistence of these intertwined orders. Therefore, the $\sigma_z$-Hubbard model proposed here may have a connection to such quantum materials, a topic we leave for further study.

\textit{Acknowledgments.---}
The authors thank M. Franz for helpful discussions. X.Z. acknowledges support from the Natural Science Foundation of Jiangsu Province under Grant BK20230907 and the NSFC grant No. 12304177. J.S. and H.G. acknowledge support from NSFC grant Nos.~11774019 and 12074022.
S.F. is supported by the National Key Research and Development 
Program of China under Grant Nos. 2023YFA1406500 and 2021YFA1401803,  
and NSFC under Grant No. 12274036.
S.S.G. is supported by the NSFC Grant No. 12274014, the Special Project in Key Areas for Universities in Guangdong Province (No. 2023ZDZX3054), and the Dongguan Key Laboratory of Artificial Intelligence Design for Advanced Materials (DKL-AIDAM). 
W. H. is supported by NSFC under Grants No. 11904155 and No. 12374042.
R.T.S. is supported by the grant DOE DE-SC0014671 funded by
the U.S. Department of Energy, Office of Science.

\appendix

\renewcommand{\thefigure}{A\arabic{figure}}

\setcounter{figure}{0}

\bibliography{ddirac}


\end{document}